# Using Linear Difference Equations to Model Nonlinear Cryptographic Sequences

**P. Caballero-Gil**

*Faculty of Mathematics, University of La Laguna, 38271 Tenerife, Spain, Email: pcaballe@ull.es*

**A. Fúster-Sabater**

*Institute of Applied Physics (CSIC), Serrano 144, 28006, Madrid, Spain, Email: amparo@iec.csic.es*

**M.E. Pazo-Robles**

*ITBA Instituto Tecnológico de Buenos Aires, Av. E Madero 399, Buenos Aires, Argentina, Email: eugepazorobles@gmail.com*

**Abstract**

A new class of linear sequence generators based on cellular automata is here introduced in order to model several nonlinear keystream generators with practical applications in symmetric cryptography. The output sequences are written as solutions of linear difference equations, and three basic properties (period, linear complexity and number of different output sequences) are analyzed.

## 1. Introduction

*Cellular Automata* (*CA*) are discrete dynamical systems of simple construction but complex behaviour. These finite state machines are defined as uniform arrays of identical cells in an n-dimensional space, and may be classified with respect to parameters such as cellular geometry, neighbourhood specifications, number of states per cell and transition rules. A well-known two-dimensional *CA* invented by Conway and popularized by Gardner[6] is the so-called Game of Life. Different mathematical techniques have been used to analyze *CA* by Wolfram, Martin and Odlyzko[15, 16, 17].

In this work, only one-dimensional binary *CA* with three site neighbourhood and linear transition rules will be used. Furthermore, the *CA* here considered are hybrid (different cells evolve under different transition rules) and null (cells with null content are adjacent to the extreme cells).

In symmetric key cryptography, most keystream generators are based on *Linear Feedback Shift Registers (LFSRs)* whose output sequences are combined in a nonlinear way. A thorough introduction to the theory of shift register sequences may be found in the classic book by Golomb[7]. The relationship between the linear *CA* above characterized and *LFSRs* was analyzed by Serra et al[13], who proved that both structures are isomorphic, and consequently the latter ones may be substituted by the former ones in order to accomplish the same goal: the generation of pseudorandom sequences. Nevertheless, as it will be shown within this paper, the main advantage of this type of *CA* is that certain multiple generators designed in terms of *LFSRs* as nonlinear structures preserve their original linearity when they are expressed under the form of linear *CA*. In particular, in this paper cryptographic generators such as the *Shrinking Generator* introduced by Coppersmith, Krawczyk and Mansour[3], and *Clock-Controlled Shrinking Generators (CCSGs)* defined by Kanso[8] will be linearly modelled through an extremely simple procedure. Furthermore, the same simple procedure can be applied to keystream generators in a wider range of practical application.

This work is organized as follows. In the next section, a new type of linear *CA* called *Multiplicative-Polynomial Cellular Automata* (*MPCA*) is introduced. Structural properties of *MPCA* are studied in section 3, where emphasis is given on different parameters of their generated sequences (e.g. period, linear complexity, characteristic polynomial and number of different output sequences). Section 4 shows that *MPCA* allow modelling nonlinear cryptographic generators in terms of linear structures. Finally, some illustrative examples and conclusions complete the paper.

## 2. Fundamentals and Basic Notation

In this section, several characteristics of the two basic structures considered within this paper (*LFSRs* and one-dimensional linear hybrid *CA*) are briefly introduced.

### 2.1. Linear Recurrence Relationship in *LFSRs*

A binary *LFSR* is a pseudorandom sequence generator made out of $L$ memory cells or stages (numbered 1, 2,..., $L$) so that each one is capable of storing one bit. At each unit of time, the following operations are performed:

(i) The content of stage 1 is output.

(ii) The content of stage $i$ is moved to stage $i-1$

$\forall i: 2 \leq i \leq L$.

(iii) The new content of stage $L$ is the exclusive-*OR* of a subset of stages given by *P(X)*, which is the *LFSR* characteristic polynomial of degree $L$. If *P(X)* is a primitive polynomial[11], then the *LFSR* is called maximal-length *LFSR* and its output sequence is a *PN*-sequence. In the sequel, only maximal-length *LFSRs* and their corresponding *PN*-sequences will be considered.

Let $\{x_n\}$ be the *PN*-sequence generated by a maximal-length *LFSR*. The linear $L$-degree recurrence relationship that specifies its $n$-th element can be written as:

$$x_n + \sum_{i=1}^{L} c_i x_{n-i} = 0, \quad n \geq L \quad (1)$$

where the sequence elements $x_n$ as well as the coefficients $c_i$ belong to *GF(2)*. In this paper both addition and multiplication refer always to *modulo* 2 operations. The linear recursion in (1) can be expressed as a linear difference equation:

$$(E^L + \sum_{i=1}^{L} c_i E^{L-i}) x_n = 0, \quad n \geq 0 \quad (2)$$

where $E$ is the shift operator defined on $x_n$ so that $Ex_n = x_{n+1}$. The characteristic polynomial of equation (2) coincides with the *LFSR* characteristic polynomial, which is:

$$P(X) = X^L + \sum_{i=1}^{L} c_i X^{L-i}. \quad (3)$$

Let $\alpha \in GF(2^L)$ be a root of *P(X)*. If *P(X)* is a primitive polynomial, then its $L$ roots are[12]:

$$\alpha, \alpha^2, \alpha^{2^2}, \ldots, \alpha^{2^{L-1}} \quad (4)$$

each of them being a primitive element of $GF(2^L)$. Thus, the $n$-th element of $\{x_n\}$ can be written in terms of the previous roots[11] such as follows:

$$x_n = \sum_{j=0}^{L-1} A^{2^j} \alpha^{2^j n} \quad (5)$$

where $A \in GF(2^L)$. The value of $A$ determines the starting point of the *PN*-sequence. It is remarkable that equation (5) is just a solution of the difference equation (2).

## 2.2. One-Dimensional Linear Hybrid *CA*

In this paper our attention is focussed on three-dimensional binary linear hybrid *CA* with three site neighbourhood. In fact, there are eight of such transition rules, among which only two (rule 90 and rule 150) lead to non trivial structures. These rules can be defined as follows:

| Rule 90 | Rule 150 |
|---|---|
| $a_{n+1}^k = a_n^{k-1} + a_n^{k+1}$ | $a_{n+1}^k = a_n^{k-1} + a_n^k + a_n^{k+1}$ |

Indeed, the content $a_{n+1}^k$ of the $k$-th cell at time $n+1$ depends on the content of either two different cells (rule 90) or three different cells (rule 150) at time $n$, $\forall k = 1, 2, \ldots, L$, where $L$ is the length of the automaton. Moreover, the state of the automaton is formed by the binary content of the $L$ cells at each unit of time. For the previous rules, the different states of the automaton are grouped in closed cycles[9]. A natural form of *CA* representation is given by an $L$-tuple $D_L= (d_1, d_2,..., d_L)$ where $d_k= 0$ if the $k$-th cell follows the rule 90 while $d_k= 1$ if the $k$-th cell follows the rule 150. Also $D_k= (d_1, d_2,..., d_k)$ $\forall k= 1, 2,..., L$ denote the corresponding sub-automata of length $k$.

| Characteristic polynomial $P(X)=X^3+X^2+1$ | | | | | | |
|---|---|---|---|---|---|---|
| LFSR | CA: 150 | 90 | 90 | 90 | 90 | 150 |
| 1 1 0 | **1** | 0 | 0 | **1** | 1 | 1 |
| 1 0 1 | **1** | 1 | 0 | **1** | 0 | 0 |
| 0 1 0 | **0** | 1 | 1 | **0** | 1 | 0 |
| 1 0 0 | **1** | 1 | 1 | **1** | 0 | 1 |
| 0 0 1 | **0** | 0 | 1 | **0** | 0 | 1 |
| 0 1 1 | **0** | 1 | 0 | **0** | 1 | 1 |
| 1 1 1 | **1** | 0 | 1 | **1** | 1 | 0 |

Tab. 1: Equal output sequences of *LFSR* and *CA*

Given an irreducible polynomial $Q(X)$, the *Cattell and Muzio synthesis algorithm*[1] provides a pair of reversal linear 90/150 *CA* whose characteristic polynomial is $Q(X)$. Reciprocally, given a linear 90/150 cellular automaton, the Euclid's *GCD* algorithm is the basis for the calculation of its corresponding characteristic polynomial. Furthermore, it is known that a linear *CA* and a *LFSR* with the same primitive characteristic polynomials are isomorphic[14]. Therefore, a one-dimensional binary linear 90/150 cellular automaton of primitive characteristic polynomial $P(X)$ given by (3) will generate the *PN*-sequence defined in equations (1) and (5). As an example, Tab. 1 depicts the same *PN*-sequence (in bold at the most left cells) generated by two different kinds of structures (*LFSR* and linear 90/150 *CA*) both with characteristic polynomial $P(X)= X^3+X^2+1$. In such an example, the *LFSR* initial sate is (1, 1, 0), while the initial states of the two reversal *CA* are (1, 0, 0) and (1, 1, 1), respectively. At the remaining cells, shifted versions of the same *PN*-sequence are generated. In the following definition, a special class of *CA* is introduced.

**Definition 1:** *A Multiplicative-Polynomial Cellular Automaton is defined as a cellular automaton whose characteristic polynomial is a reducible polynomial of the form $P_M(X)= P(X)^p$ where p is a positive integer and P(X) is an irreducible polynomial. If P(X) is a primitive polynomial, then the automaton is called a Primitive Multiplicative-Polynomial Cellular Automaton (PMPCA).*

The polynomial $P_M(X)$ is a reducible polynomial, so the *Cattell and Muzio algorithm* can not be applied. Nevertheless, in the next section, linear 90/150 *CA* with characteristic polynomials $P_M(X)$ are introduced.

## 3. Properties of *MPCA*

Since the characteristic polynomial of *MPCA* is of the form $P_M(X)= P(X)^p$, it seems quite natural to construct a *Multiplicative-Polynomial Cellular*

*Automaton* by concatenating *p* times the basic *CA* of characteristic polynomial *P(X)*. The following lemma is a concrete formalization of this idea.

**Lemma 2:** *Let O be a linear hybrid 90/150 cellular automaton of length L, binary codification ($d_1, d_2,..., d_{L-1}, d_L$) and characteristic polynomial P(X). Let Õ be the reversal version of O, with binary codification ($d_L, d_{L-1},..., d_2, d_1$), and the same length and polynomial as O. Then, the 2L-tuple $(d_1, d_2, ..., \overline{d_L}, \overline{d_L}, ..., d_2, d_1)$ represents the linear 90/150 cellular automaton of length 2L and characteristic polynomial $P(X)^2$.*

The proof is based on the recurrence relationship for the characteristic polynomials of the successive sub-automata of a given automaton[1]. Let $\Delta_k(X)$ denote the characteristic polynomial of the sub-automaton ($d_1, d_2, ..., d_k$) and let $\Delta_k(X) = (X + d_k)\Delta_{k-1}(X) + \Delta_{k-2}(X)$ ($k>0$, $\Delta_{-1}=0$, $\Delta_0=1$) be the above mentioned recurrence relationship. Then, the successive polynomials of the previous 2L-tuple are:

$$\Delta_{2L} = (X + d_1)\Delta_{2L-1} + \Delta_{2L-2}$$
$$\Delta_{2L-1} = (X + d_2)\Delta_{2L-2} + \Delta_{2L-3}$$
$$\Delta_{2L-2} = (X + d_3)\Delta_{2L-3} + \Delta_{2L-4}$$
$$\vdots$$
$$\Delta_{L+2} = (X + d_{L-1})\Delta_{L+1} + \Delta_L$$
$$\Delta_{L+1} = (X + \overline{d_L})\Delta_L + \Delta_{L-1}$$
$$\Delta_L = (X + \overline{d_L})\Delta_{L-1} + \Delta_{L-2}$$
$$\Delta_{L-1} = (X + d_{L-1})\Delta_{L-2} + \Delta_{L-3}$$
$$\vdots$$
$$\Delta_3 = (X + d_3)\Delta_2 + \Delta_1$$
$$\Delta_2 = (X + d_2)\Delta_1 + 1$$
$$\Delta_1 = (X + d_1)$$

Thus, the computation of $\Delta_{2L}$ can be carried out by multiple substitutions.

$$\Delta_{2L} = \Delta_1\Delta_{2L-1} + \Delta_{2L-2} = \Delta_2\Delta_{2L-2} +$$
$$+ \Delta_1\Delta_{2L-3} = \Delta_3\Delta_{2L-3} + \Delta_2\Delta_{2L-4} = \cdots =$$
$$= \Delta_{L+1}\Delta_{L-1} + \Delta_L\Delta_{L-2} = (X + \overline{d_L})\Delta_L\Delta_{L-1} +$$
$$+ \Delta_{L-1}^2 + (X + \overline{d_L})\Delta_{L-1}\Delta_{L-2} + \Delta_{L-2}^2 =$$
$$= (X + \overline{d_L})\Delta_{L-1}(\Delta_L + \Delta_{L-2}) + \Delta_{L-1}^2 +$$
$$+ \Delta_{L-2}^2 = (X + \overline{d_L})\Delta_{L-1}(X + \overline{d_L})\Delta_{L-1} +$$
$$+ \Delta_{L-1}^2 + \Delta_{L-2}^2 = (X + \overline{d_L})^2\Delta_{L-1}^2 + \Delta_{L-2}^2 =$$
$$= ((X + d_L)\Delta_{L-1} + \Delta_{L-2})^2.$$

If *P(X)* is the characteristic polynomial of the automaton ($d_1, d_2, ..., d_L$), then the characteristic polynomial of the automaton $(d_1, d_2, ..., \overline{d_L}, \overline{d_L}, ..., d_2, d_1)$ is:

$$\Delta_{2L}(X) = P(X)^2. \tag{6}$$

The basic automaton is concatenated with its reversal version after the complementation of the last rule $d_L$. Consequently, successive applications of this result provide *MPCA* whose characteristic polynomials are: $P(X)^2, P(X)^{2^2}, ..., P(X)^{2^q}, ...$ of lengths $2L, 2^2L, ..., 2^qL, ...,$ respectively. It is remarkable that for every *P(X)* there are two different basic automata that may be used for the concatenation. Therefore, if $2^{q-1} < p \leq 2^q$, then the two *MPCA* of length $2^q L$ built as in *Lemma 2* and applied on different initial states will produce all the sequences $\{a_n\}$ with characteristic polynomial $P(X)^p$ that satisfy the difference equation:

$$(E^L + \sum_{i=1}^{L} c_i E^{L-i})^p a_n = 0, \quad n \geq 0 \qquad (7)$$

On the other hand, if $P_M(X) = P(X)^p$, then the roots of $P_M(X)$ will be the same as those of $P(X)$ but with multiplicity $p$. Thus, the *n*-th element of $\{a_n\}$ can be written in terms of the previous multiple roots such as follows:

$$a_n = \sum_{j=0}^{L-1} \sum_{i=0}^{p-1} (n_i A_i^{2^j}) \cdot \alpha^{2^j n} \qquad (8)$$

where $A_i \in GF(2^L)$ and $n_i$ are binomial coefficients reduced *modulo* 2. The choice of $A_i$ determines the properties of the sequences $\{a_n\}$ generated by *MPCA*.

### 3.1. Period of Sequences Generated by MPCA

The solutions of the equation (7) can be rewritten as:

$$a_n = \sum_{i=0}^{p-1} (n_i \sum_{j=0}^{L-1} A_i^{2^j} \alpha^{2^j n}). \qquad (9)$$

According to equation (5), $x_n^i = \sum_{j=0}^{L-1} A_i^{2^j} \alpha^{2^j n}$ represents the *n*-th element of the *PN*-sequence of period $2^L - 1$ whose starting point is determined by $A_i$. Thus, $\{a_n\}$ can be written as the sum of *p* times the same *PN*-sequence starting at different points and weighted by binomial coefficients

$$\{a_n\} = \sum_{i=0}^{p-1} n_i \{x_n^i\}. \qquad (10)$$

In addition, each binomial coefficient defines a succession of binary values of constant period $p_i$. Tab. 2 shows the values of $p_i$ for different $n_i$. Such as it may be seen, they are different powers of 2 between the integers 1 and *p*. Therefore, the sequence $\{a_n\}$ is the sum of *p* sequences of distinct periods $T_i = p_i \cdot (2^L - 1)$, and the period of such a sum sequence will be:

$$T = max\{T_i \ (i = 0,..., p - 1) \ / \ A_i \neq 0\}. \qquad (11)$$

It can be noticed that the period of the different sequences $\{a_n\}$ generated by an *MPCA* depends on the choice of the coefficients $A_i$ in equation (8). Nevertheless, all the sequences generated at the same state cycle have the same period.

| Binomial coeff. | Binary values | $p_i$ |
|---|---|---|
| $n_0$ | 1,1,1,1,1,1,1,1,1,1,… | $p_0=1$ |
| $n_1$ | 0,1,0,1,0,1,0,1,0,1,… | $p_1=2$ |
| $n_2$ | 0,0,1,1,0,0,1,1,0,0,… | $p_2=4$ |
| $n_3$ | 0,0,0,1,0,0,0,1,0,0,… | $p_3=4$ |
| $n_4$ | 0,0,0,0,1,1,1,1,0,0,… | $p_4=8$ |
| $n_5$ | 0,0,0,0,0,1,0,1,0,0,… | $p_5=8$ |
| $n_6$ | 0,0,0,0,0,0,1,1,0,0,… | $p_6=8$ |
| $n_7$ | 0,0,0,0,0,0,0,1,0,0,… | $p_7=8$ |

Tab. 2: Coefficients, values and periods

### 3.2. Linear Complexity of Sequences Generated by *MPCA*

The linear complexity of a sequence equals the number of roots (with their multiplicities) that appear in the linear recurrence relationship[10]. Therefore, going back to expression (8), the linear complexity of $\{a_n\}$ can be computed because there are $L$ roots each of them with multiplicity $p$. Thus, if $i_{max}$ is the greatest value of $i$ ($i= 0,1,...,p-1$) for which $A_i \neq 0$, then the *linear complexity LC* of the sequence $\{a_n\}$ will be:

$$LC = (i_{max} + 1) \cdot L. \qquad (12)$$

The maximum linear complexity will be $LC_{max}= p \cdot L$ (if $A_{p-1}\neq 0$) while the minimum linear complexity will be $LC_{min}= L$ (if $A_i=0\ \forall i>0$). The linear complexity for this kind of sequences will always be a multiple of $L$. In brief, the linear complexity of the different sequences $\{a_n\}$ depends on the choice of the coefficients $A_i$ in (8), and all the sequences generated at the same state cycle have the same linear complexity.

### 3.3 Number of Different Sequences Generated by *MPCA*

In order to get the number of different sequences $\{a_n\}$ generated by *MPCA*, the choice of the coefficients $A_i$ in equation (8) must be considered. Three distinct situations may be distinguished:

• If $A_i= 0\ \forall i$, then all the cells of the *CA* will generate the identically null sequence.

• If $A_0\neq 0$ and $A_i= 0\ \forall i> 0$, then all the cells of the *CA* will generate a unique *PN*-sequence $\{x_n\}$ of period $T_0= 2^L-1$ and characteristic polynomial $P(X)$. It is remarkable that the relative shifts of this sequence generated at the different cells can be determined[2].

• In general, if $A_0,A_1,...,A_{i-1}\in GF(2^L)$, $A_i\neq 0$ and $A_j= 0\ \forall j>i$, $i\geq 1$, then there are $2^{iL} \cdot (2^L-1)$ possible choices of $(A_0,A_1,...,A_i)$. According to section 3.1, the period of such sequences is the maximum value of $T_i= p_i \cdot (2^L-1)$. Thus, the number of different sequences for these values of $A_i$ is:

$$N_i = \frac{2^{iL} \cdot (2^L-1)}{T_i} = \frac{2^{iL}}{p_i}. \qquad (13)$$

Consequently, the total number of distinct sequences obtained from an *MPCA* (excluded the null sequence) is:

$$N_{total} = \sum_{i=0}^{p-1} N_i. \qquad (14)$$

### 3.4. Illustrative Example

This section includes a simple example to illustrate the previous results. Consider the *PMPCA* of 20 cells $D_{20}=$ (1,0,0,0,1,1,0,0,0,0,0,0,0,1,1,0,0,0,1) with characteristic polynomial $P_M(X)= P(X)^p$, $P(X)= X^5+ X^4+ X^2+ X+ 1$, $p= 4$, $L= 5$. Different choices of $A_i$ (not all null) are now considered separately:

1. If $A_0\neq 0$ and $A_i= 0\ \forall i> 0$, then the *CA* will produce $N_0= 1$ sequence, which is a unique *PN*-sequence of period $T_0= 31$, linear complexity $LC_0 = 5$ and characteristic polynomial $P(X)$. In addition, the automaton cycles through doubly symmetric states of the form: $(a_0, a_1, a_2, a_3, a_4, a_4, a_3, a_2, a_1, a_0, a_0, a_1, a_2, a_3, a_4, a_4, a_3, a_2, a_1, a_0)$ with $a_i \in GF(2)$. Fig. 1 illustrates the formation of the output sequences (binary sequences in vertical) for the previous $CA$ of 20 cells and initial state (1,1,0,0,1,1,0,0,1,1,1,1,0,0,1,1,0,0,1,1). In fact, diamonds represent 1's and blanks represent 0's. The 31 doubly symmetric states are concentrated into the same cycle.

2. If $A_0 \in GF(2^5)$, $A_1 \neq 0$ and $A_i = 0$ $\forall i > 1$, then the $CA$ will produce $N_1 = 16$ different sequences of period $T_1 = 62$, linear complexity $LC_1 = 10$ and characteristic polynomial $P(X)^2$. Moreover, the automaton cycles through symmetric states of the form: $(a_0, a_1, a_2, a_3, a_4, a_5, a_6, a_7, a_8, a_9, a_9, a_8, a_7, a_6, a_5, a_4, a_3, a_2, a_1, a_0)$ with $a_i \in GF(2)$. Fig. 2 illustrates the formation of the output sequences for the previous cellular automaton of 20 cells and initial state (0,0,0,0,0,0,0,0,0,1,1,0,0,0,0,0,0,0,0,0). In fact, there are $2^{10} - 32 = 992$ symmetric states distributed in 16 cycles of 62 states each of them.

3. If $A_0, A_1 \in GF(2^5)$, $A_2 \neq 0$ and $A_i = 0$ $\forall i > 2$, then the $CA$ will produce $N_2 = 256$ different sequences of period $T_2 = 124$, linear complexity $LC_2 = 15$ and characteristic polynomial $P(X)^3$. Moreover, the automaton cycles through several repetitive states of the form: $(a_0, a_1, a_2, a_3, a_4, a_5, a_6, a_7, a_8, a_9, a_0, a_1, a_2, a_3, a_4, a_5, a_6, a_7, a_8, a_9)$ with $a_i \in GF(2)$.

4. If $A_0, A_1, A_2 \in GF(2^5)$, $A_3 \neq 0$, then the cellular automaton will produce $N_3 = 8192$ different sequences of period $T_3 = 124$, linear complexity $LC_3 = 20$ and characteristic polynomial $P(X)^4$. In addition, the automaton cycles through the states not included in the previous cycles.

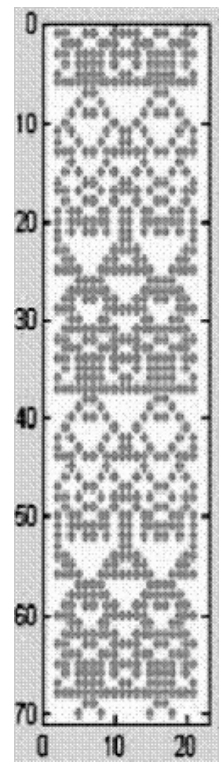

Fig. 1: $CA$ $8C031_{Hex}$ with initial state $CCF33_{Hex}$

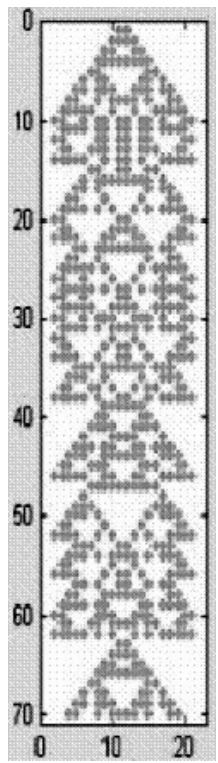

Fig. 2: *CA* 8C031$_{Hex}$ with initial state CCF33$_{Hex}$

## 4. *MPCA*-Based Model of Cryptographic Generators

The previous analysis of *MPCA* can be used for the linearization of cryptographic generators. In particular, the *Shrinking Generator* and the class of *CCSGs* are typical examples of binary sequence generators with practical application in symmetric cryptography. These generators are based on two *LFSRs* where the output bits of one register decimate the sequence produced by the other. The resultant decimated sequence is just the output sequence of the generator. The properties of these generators can be summarized as follows:

• The *shrinking generator*[3] is made of two *LFSRs*, *SR1* and *SR2*, with lengths $L_j$ (*j*= 1, 2) and characteristic polynomials $P_j(X)$ (*j*= 1, 2) respectively. The decimation rule is: The bit produced by *SR2* is discarded if the corresponding bit of *SR1* equals 0. The period of the generated sequence is $T = (2^{L_2} - 1)2^{L_1 - 1}$ and its linear complexity takes values in the interval $L_2 2^{L_1 - 2} < LC \leq L_2 2^{L_1 - 1}$. The characteristic polynomial is of the form $P_M(X) = P(X)^p$, $P(X)$ being a primitive polynomial of degree $L = L_2$ and $2^{L_1 - 2} < p \leq 2^{L_1 - 1}$. Moreover, $P(X)$ is the characteristic polynomial[4,5] of the cyclotomic coset $E$ in $GF(2^{L_2})$ with $E = 2^0 + 2^1 + \cdots + 2^{L_1 - 1}$.

• A Clock-Controlled Shrinking Generator[8] is made out of two *LFSRs*, *SR1* and *SR2*, with lengths $L_j$ (*j*= 1, 2) and characteristic polynomials $P_j(X)$ (*j*= 1, 2), respectively, plus a decimation function $DF_t$ that depends on the bits of *SR1* at each unit of time. Period and linear complexity are analogous to those of the previous generator. Indeed, a *CCSG* is a generalized version of the *shrinking generator*. So, the characteristic polynomial is of the form $P_M(X) = P(X)^p$, $P(X)$ being a primitive polynomial of degree $L = L_2$ that depends on $P_2(X)$, $L_1$ and the decimation function $DF_t$. In addition, *p* takes values in the same interval as before.

Both classes of generators produce sequences with characteristic polynomials of the form $P_M(X) = P(X)^p$. Thus, their output sequences will be solutions of linear difference equations corresponding to *PMPCA*. Consequently, these generators can be expressed in terms of a linear model based on *CA*. A simple example for the *shrinking generator* illustrates the simple modelling procedure:

***Input:*** A *shrinking generator* characterized by two *LFSRs* of lengths $L_1 = 3$ and $L_2 = 5$ and characteristic polynomial $P_2(X) = X^5 + X^4 + X^2 + X + 1$.

• *Step 1:* $P(X)$ is the characteristic polynomial of the cyclotomic coset $E = (2^0 + 2^1 + 2^2)$ in $GF(2^5)$. Thus, the polynomial $P(X)$ is of degree $L = L_2 = 5$:

$$P(X) = (X + \alpha^E)(X + \alpha^{2E}) \cdots (X + \alpha^{2^{L-1}E}) =$$
$$= X^5 + X^2 + 1.$$

• *Step 2:* Applying the *Cattell and Muzio algorithm*[1], two linear *CA* whose characteristic polynomial is $P(X)$ can be determined. Such *CA* are written in binary codification as:

0 1 1 1 1

1 1 1 1 0

• *Step 3:* Computation of the required pair of *CA* by successive concatenations.

For the first automaton:

0 1 1 1 1

0 1 1 1 0 0 1 1 1 0

0 1 1 1 0 0 1 1 1 1 1 1 1 0 0 1 1 1 0

For the second automaton:

1 1 1 1 0

1 1 1 1 1 1 1 1 1

1 1 1 1 1 1 1 1 0 0 1 1 1 1 1 1 1 1 1

For each automaton, the procedure in *Step 3* has been carried out $L_1 - 1$ times. In fact, each basic automaton with complementations has been concatenated $p = 2^{L_1 - 1} = 4$ times.

***Output:*** Two binary strings of length $p \cdot L = 2^{L_1 - 1} \cdot L_2$ corresponding to the *CA*.

In this way, we have obtained a pair of linear *CA* able to produce the shrunken sequence corresponding to the given *shrinking generator*. An analogous procedure may be applied for a *CCSG*. In brief, we have obtained two simple and different linear models describing the behaviour of a nonlinear cryptographic sequence generator.

## 5. Conclusions

In this work, a new type of *CA* called *Primitive Multiplicative-Polynomial Cellular Automata* have been introduced and analyzed. What it is important about *PMPCA* is that it has been shown that a wide class of *LFSR*-based sequence generators of practical cryptographic application, such as the *Shrinking Generator* and the *Clock-Controlled Shrinking Generators*, can be described in terms of *PMPCA*-based structures. In this way, sequence generators conceived and designed as complex nonlinear models can be written in terms of simple linear models. Furthermore, the algorithm to convert a given nonlinear *LFSR*-based generator into a linear *CA*-based model is very simple and can be applied to generators in a wide range of practical interest. Thus, the linearity of these cellular models might be advantageously used in the analysis and/or cryptanalysis of such keystream generators.

## Acknowledgements

This work was supported in part by Ministry of Science and Innovation and European FEDER Fund under Project TIN2008-02236/TSI, as well as by CDTI (Spain) and the companies INDRA, Unión Fenosa, Tecnobit, Visual Tool, Brainstorm, SAC and Technosafe under Project Cenit-HESPERIA.